\documentclass[prd,twocolumn]{revtex4}
\usepackage{graphicx}
\usepackage{amssymb}
\usepackage{epstopdf}

\begin{document}

\title{Finding the Missing Baryons Using CMB as a Backlight}
\author{Shirley Ho}
\email{cwho@lbl.gov}
\affiliation{Department of Astrophysical Sciences, Princeton University, Princeton, NJ 08544 \\ \& \\ Lawrence Berkeley National Laboratory, Berkeley, CA 94704}
\author{Simon DeDeo}
\affiliation{Kavli Institute for Cosmological Physics, University of Chicago, Chicago, IL 60637, USA \\ \& \\ Institute for the Physics and Mathematics of the Universe, University of Tokyo, Kashiwano-ha 5-1-5, Kashiwa-shi, Chiba 277-8582, Japan}
\author{David N. Spergel}
\affiliation{Department of Astrophysical Sciences, Princeton University, Princeton, NJ 08544~}
\date{\today}

\begin{abstract}
We present a new method for detecting the missing baryons by generating a template for the kinematic Sunyaev-Zel'dovich effect.
The template is computed from the product of a reconstructed velocity field
with a galaxy field; we find that the combination of a galaxy redshift survey such as SDSS and
a CMB survey such as ACT and PLANCK can detect the kSZ, and thus the ionized gas, at significant signal-to-noise.
Unlike other techniques that look for hot gas or metals, this approach directly detects the
electrons in the IGM through their signature on the CMB.
The estimated signal-to-noise for detecting the galaxy-momentum kSZ cross-correlation
is 4, 9, and 12 for ACT (with survey area of 2000 $\mathrm{deg}^2$) with SDSS-DR4, SDSS3 and ADEPT respectively.
The estimated signal-to-noise  for PLANCK with SDSS-DR4, SDSS3 and ADEPT is 11, 23, and 32.
Our method provides a new mean for determining properties of the ionized gas in the Universe. We provide galaxy momentum templates constructed from Sloan Digital Sky Survey online at 
our website at {\tt http://www.astro.princeton.edu/$\sim$shirley/SZ/SZ.html}. 
The predicted correlation coefficients are provided along with the momentum maps.
One can download the momentum templates and cross-correlate directly with CMB maps from ACT and PLANCK to detect
the missing baryons.
\end{abstract}

\maketitle
Where are the baryons?  Astronomers can measure the baryon abundance three minutes after the big bang using
Deuterium~\cite{gs07}, 300,000 years after the big bang using the angular power spectrum of
the Cosmic Microwave Background~\cite{s07}, and at redshift three using the Lyman-$\alpha$ forest~\cite{r98}.
However, today, most of the baryons are missing~\cite{b07}.  Too hot to show up in Lyman-$\alpha$ absorption studies,
too cool to find in spectral distortions of the cosmic microwave background, and hidden
in large-scale structures of the cosmic web of too low a density to appear in the X-ray,
finding this missing matter is one of the open challenges in cosmology.

The baryons within clusters -- shock heated during virialization of the structure --
have been the subject of extensive study with X-ray telescopes \citep{forman82,sarazin86,rosati02}.
However, as X-ray emission strength is proportional to the square of electron density, these studies can only detect baryons in the very center of clusters. There have been numerous
attempts to make X-ray observations of the filamentary gas \citep{shull05},
but filamentary gas seems to elude X-ray detection.

Baryons within clusters are now being detected at significant confidence levels through their distortion of the
primordial black-body spectrum, both in observations targeted towards known clusters~\cite{g04}, in
cross-correlation with the WMAP data~\cite{c07,h04,dunkley08}, and now in blind surveys as the latest detectors come online~\cite{jeff08}. This effect, known as thermal Sunyaev-Zel'dovich (hereafter tSZ, \cite{sz80}),
is, at the arcminute scales to be probed by current and upcoming experiments such as the \emph{Atacama Cosmology Telescope} (ACT),
the \emph{South Pole Telescope} (SPT) and the \emph{Planck} satellite,  the strongest signal (in two-point correlation) on the microwave sky.
Measurements of the tSZ effect can detect baryons in clusters. The tSZ, however, depends for its strength on
both the density and the temperature of the gas, and so an experiment that may detect the gas at the center of a cluster,
where temperatures can reach over $10^7$ K, will be unable to find at all the gas associated with the far less dense, and far colder, material that lies in the filaments of the cosmic web~\cite{ms04}.

Even when a survey reaches sensitivities of 2 $\mu\mathrm{K}~\mathrm{arcmin}^{-2}$,
identifiable tSZ sources are all objects with mass greater than $2\times 10^{14} M_\odot$~\cite{sehgal07},
and most of the contribution to the power spectrum comes from virialized gas in groups just below the
detection mass~\cite{komatsu02}.

%However, according to simulations, most of the baryons lie in the filaments between galaxies~\cite{co06}. 
The kinetic Sunyaev-Zel'dovich effect~\cite[hereafter kSZ]{sz80}, in contrast to the tSZ,
depends not on the thermal motions of the gas, but rather on the bulk velocity of the structures the gas inhabits.
In particular, the flows of matter toward overdensities Doppler-shift CMB photons to produce anisotropies.
The fractional change in temperature due to kSZ, $\Theta=\Delta T/T_{cmb}$, is
\begin{equation}
\label{integral}
\Theta(\hat{n})=-\int^{\eta_0}_0 ~d\eta ~g(\eta)\mathbf{n}\cdot\mathbf{p}(\hat{n}\eta,\eta),
\end{equation}
where $\mathbf{n}$ is the unit vector pointing away from the observer and the momentum field, $p$, is defined as
\begin{equation}
\label{momentum}
\mathbf{p}(\hat{n}\eta,\eta)=[1+\delta_b(\hat{n}\eta)]\mathbf{v_b}(\hat{n}\eta,\eta),
\end{equation}
and $\delta_b$ is the baryon overdensity, and the $\mathbf{v_b}$ is the baryon velocity.
The visibility function $g$ is defined as
\begin{equation}
\label{visibility}
g(\eta)=x_e\tau_H(1+z)^2e^{-\tau(z)},
\end{equation}
and $x_e$ is the ionization fraction, $\tau_H$ is the Thompson-scattering optical depth to the Hubble distance today; $\tau$, in the reionized epoch, is
\begin{equation}
\label{tau}
\tau(z) = \frac{2}{3} \frac{\tau_H}{\Omega_m}[\sqrt{1-\Omega_m + \Omega_m(1+z)^3} -1]
\end{equation}
and, finally, $\eta$ is the comoving distance defined in units of Hubble distance: 
\begin{equation}
\label{comov}
\eta(z) = \int_0^z \frac{H_0}{H(z')} dz'.
\end{equation}

Baryons in the cosmic web have sufficient velocity and column density to produce a detectable kSZ signal~\cite{dst05}.
Indeed, at arcminute scales, the kSZ is expected to be the strongest anisotropy
after subtraction of the tSZ using frequency information~\cite{dhs04}. Here we demonstrate a new method for
extracting the kSZ that is not subject to the systematics
one might expect when trying to see it in the two-point CMB auto-correlation.

In particular, we consider the idea of cross-correlating the CMB sky with a kSZ template constructed by projecting down
the line-of-sight momentum field of large scale structure. This momentum field is reconstructed from the inferred three-dimensional dark matter distribution.
We construct such a template from the latest large scale structure data, and make it available to the scientific community.

To generate the momentum field we require knowledge of the velocity field.
We can rely upon the linearity of the large-scale velocity field, $\mathbf{v}(\mathbf{k})$, which is related to the density field via the continuity equation,
\begin{equation}
\label{velocity}
\mathbf{v}(\mathbf{k})=i\frac{d\ln{G}}{d\ln{a}}\frac{aH\delta_m(\mathbf{k})\mathbf{k}}{k^2},
\end{equation}
where $a$ is the scale factor, $G$ is the growth factor at late times (proportional to $a$ in the matter dominated regime), $\delta_m$ is the matter density fluctuation field and $k$ is the comoving wave number. This is strictly valid only on larger scales where material remains in the linear regime; we shall determine how well it holds in approximation and on different scales in the following section.

The temperature increment for a particular velocity estimate depends on both the cosmological parameters
through Eq.~\ref{velocity}, and on the gas column density and thus, in conjunction
with other observations that can constrain cosmological parameters, can unambiguously
detect the missing baryons in the large, low-density filaments that have undetectable tSZ.

There are two major advantages to this method over others that have been proposed.
The first is that we expect the sign of the line-of-sight velocity to be uncorrelated with
many of the standard systematics, such as insufficiently subtracted tSZ, galactic foregrounds, and detector and telescope noise.

Secondly, our method makes near-maximal use of the information available in a potential cross-correlation.
Previous studies of constraining dark energy parameters by using kSZ information in cross-correlation with
a CMB signal have ``thrown out'' some of the signal, dropping the phase
information that determines the sign of the kSZ signal, and information about where velocity
flows are most likely to be found.

The idea of using peculiar velocity flows in a galaxy survey to do cosmology has been studied in past decades, most notably by the use of redshift and luminosity distance indicators for very nearby ($z<0.1$) objects \cite{strauss95}.
However, the method we propose here has only become technically feasible in recent years with the
combination of large scale redshift surveys and
the current and upcoming arcminute scale CMB experiment,
and the issues and questions we must confront are very different.

Our paper has four parts: the reconstruction of the velocity field (Sec.~\ref{vsection}),
the determination of the cross-correlation coefficients (Sec.~\ref{kszcl}), signal-to-noise estimates for current and forthcoming surveys (Sec.~\ref{sn})
and the template construction (Sec.~\ref{data}) using actual large scale structure data from SDSS.

\section{Velocity Field Reconstruction}

\label{vsection}
We first detail our method of velocity reconstruction, which provides us with the velocity field that
is accurate in large range of scales which are relevant to the structures that harbor most of the gas in the Universe.
We also validate our reconstruction method via a N-body simulation.

\subsection{Theory}

The Universe provides us with the redshift space overdensity field, and there are many ways we can construct
a real space density (see, \emph{e.g.}, \cite{tegmark04}). We
take here the simplest approach, since the scales we are interested in are not
strongly affected by non-linear effects -- fingers of God, for example, occur on cluster scales, whereas the velocity field has most power on scales larger than 10~Mpc.
We directly transform the redshift density field into a real space density field using the Friedman equation (with WMAP5 parameters), and embed the real space density field in a box. We compute the overdensity field in real space, convert it to Fourier space, and use the transformation described by
Eq.~\ref{velocity} to determine the velocity field. We then convert this field back into real space, and use this representation
to study the properties of cosmological velocity flows.

Our method falls into six steps.
\begin{enumerate}
\item
Transform the redshift space overdensity field to real space overdensity field.
\item
Fourier transform the overdensity field.
\item
Wiener-filter (as discussed below) the map given our knowledge of noise from Poisson statistics and non-linearities.
\item
Apply Eq.~\ref{velocity} to compute $\mathbf{v}(k)$.
\item
Inverse Fourier transform $\mathbf{v}(k)$ to find the real space velocities.
\item
Transform the real space velocities back to redshift space, and find the momentum field by taking the product with a density survey (which may or may not be the same as the survey used to determine the velocities.)
\end{enumerate}

While this method sounds simple, its error properties can be hard to estimate. In particular, even if observational noise
in the density field is uncorrelated (\emph{e.g.}, in the case of Poisson noise alone), the noise in the velocity field will be correlated from point to point. Formally, this can be seen by noting that the Fourier transform of $1/k$ is proportional to $1/r^2$, and so in real space computing the velocity field corresponds to convolving with the gravitational force kernel. The induced statistical errors remain Gaussian -- but their spatial correlation is no longer a delta function.

It is also important to recognize that points near the edges of the survey will have errors from our lack of knowledge of what lies beyond the survey volume. Intuitively, one can imagine the possibility of a ``great attractor'' lying just outside the boundary~\cite{masters07,masters08}, changing the velocity field radically from the estimate produced by using only the observed material. Developing mathematical techniques to determine the influence of these errors on the momentum field is difficult. We side-step the problem in this section by using an N-body simulation, where we have knowledge of the true velocity field.

\subsection{Simulation}
\label{vsim}

We check the validity of our reconstruction method with simulations. We first simulate a galaxy field and then implement our method of computing the velocity field. We then compare our recovered velocity field to the true velocity field in the simulation and investigate how well the reconstruction has worked.

Our N-body simulation uses WMAP 5-year parameters, has volume $256^3~h^{-1}$~Mpc and contains $256^3$ particles. The linear CDM power spectrum was generated using code from the
GRAFIC2~\footnote{Available at {\tt http://arcturus.mit.edu/grafic/}} package~\citep{Bert01}. GRAFIC2 was then used to generate the initial particle conditions, with the modification that the Hanning filter was not used because it suppresses power on small scales~\citep{MTC05}. Simulations were carried out using the TPM~\footnote{Available at {\tt http://www.astro.princeton.edu/$\sim$bode/TPM/}} code~\citep{BO03} with a $256^3$ mesh and a spline softening length of 20.35$~h^{-1}$kpc.  The initial domain decomposition
parameters in the TPM code were $A=1.9$ and $B=8.0$ (TPM was modified slightly so that there was no lower limit to $B$ when it is reduced at later times, which improves the tracking of low mass halos; for details on these parameters see \citep{BO03}).

\begin{figure}
\includegraphics[width=3.0in]{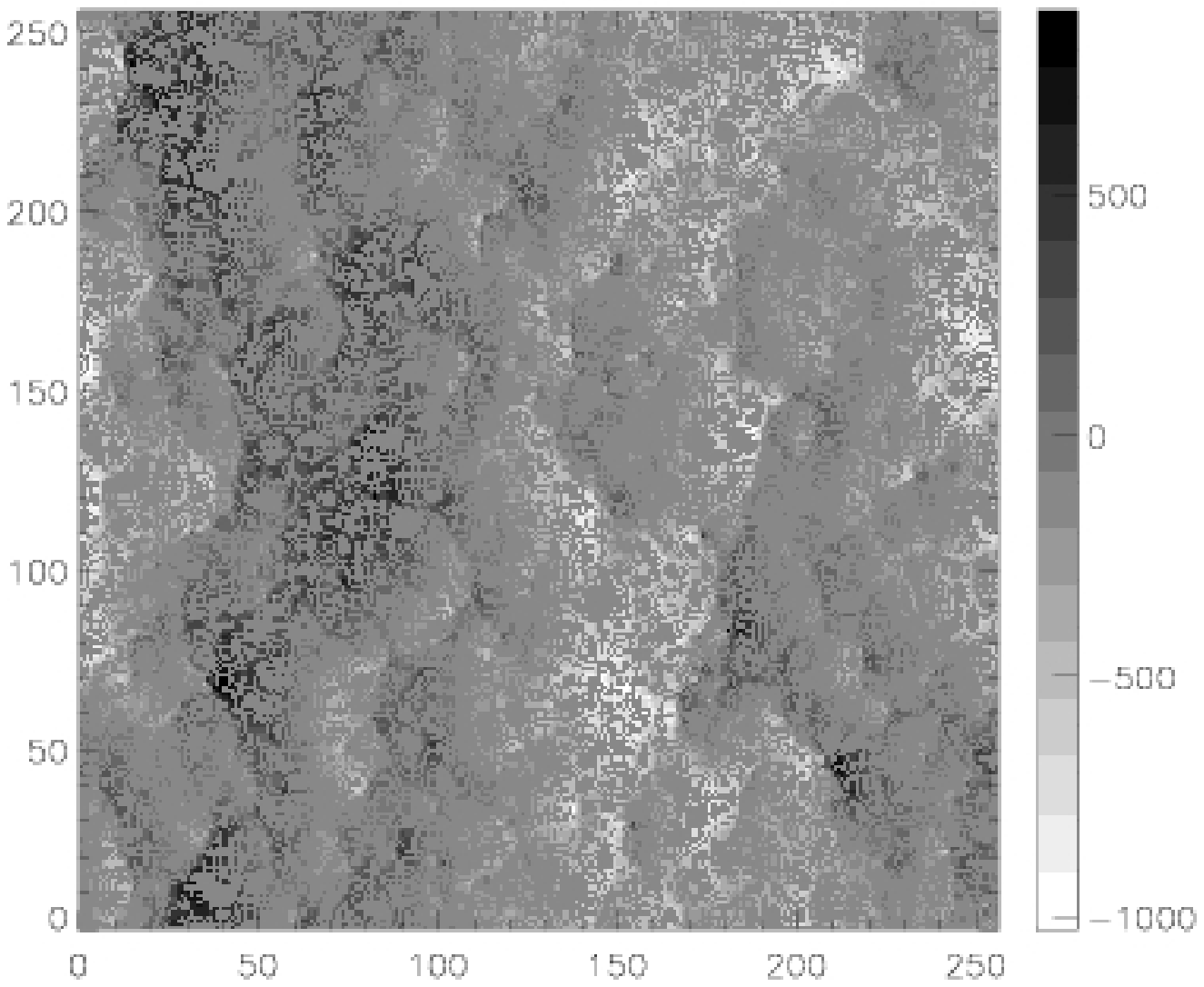}
\includegraphics[width=3.0in]{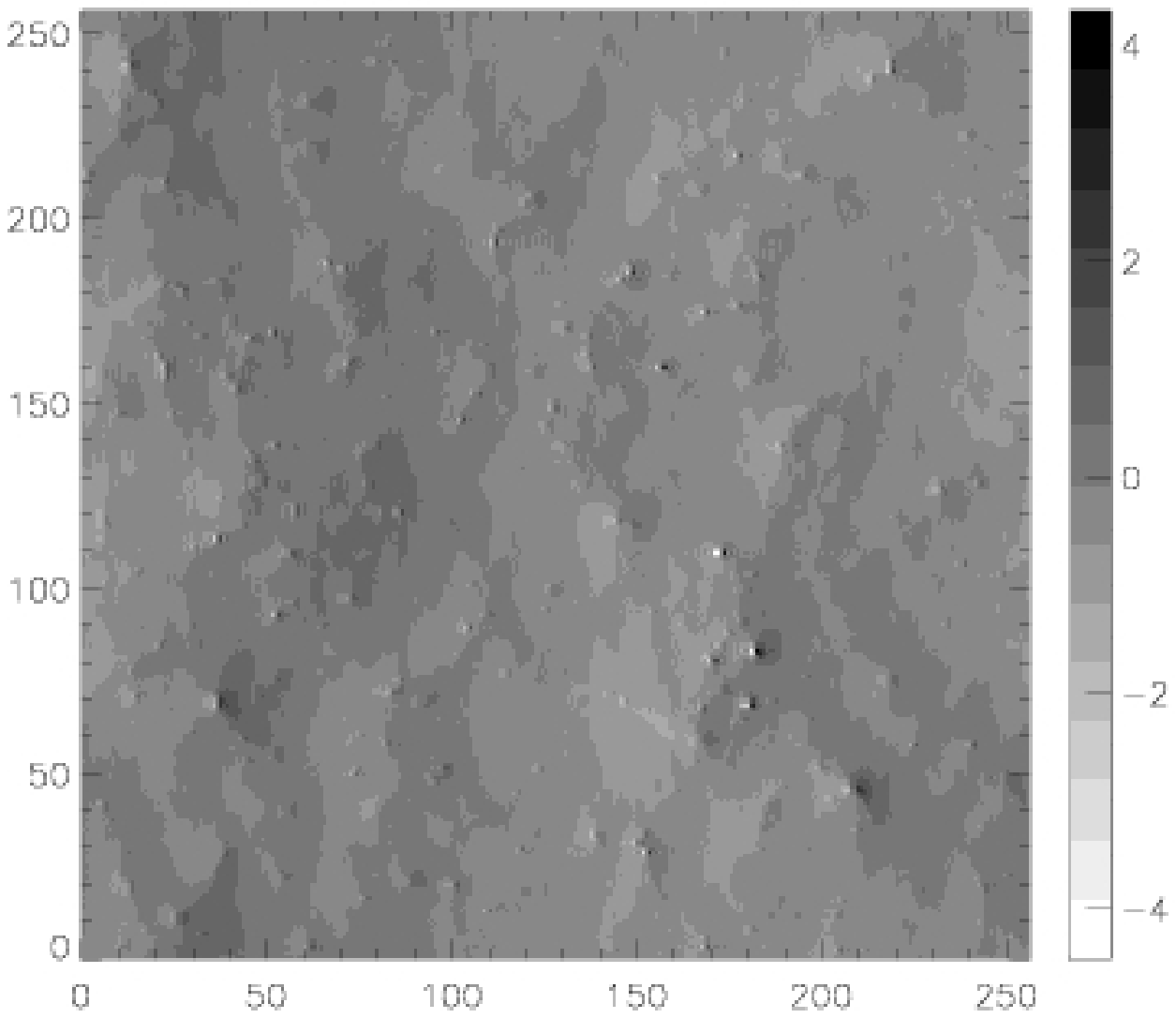}
\includegraphics[width=3.0in]{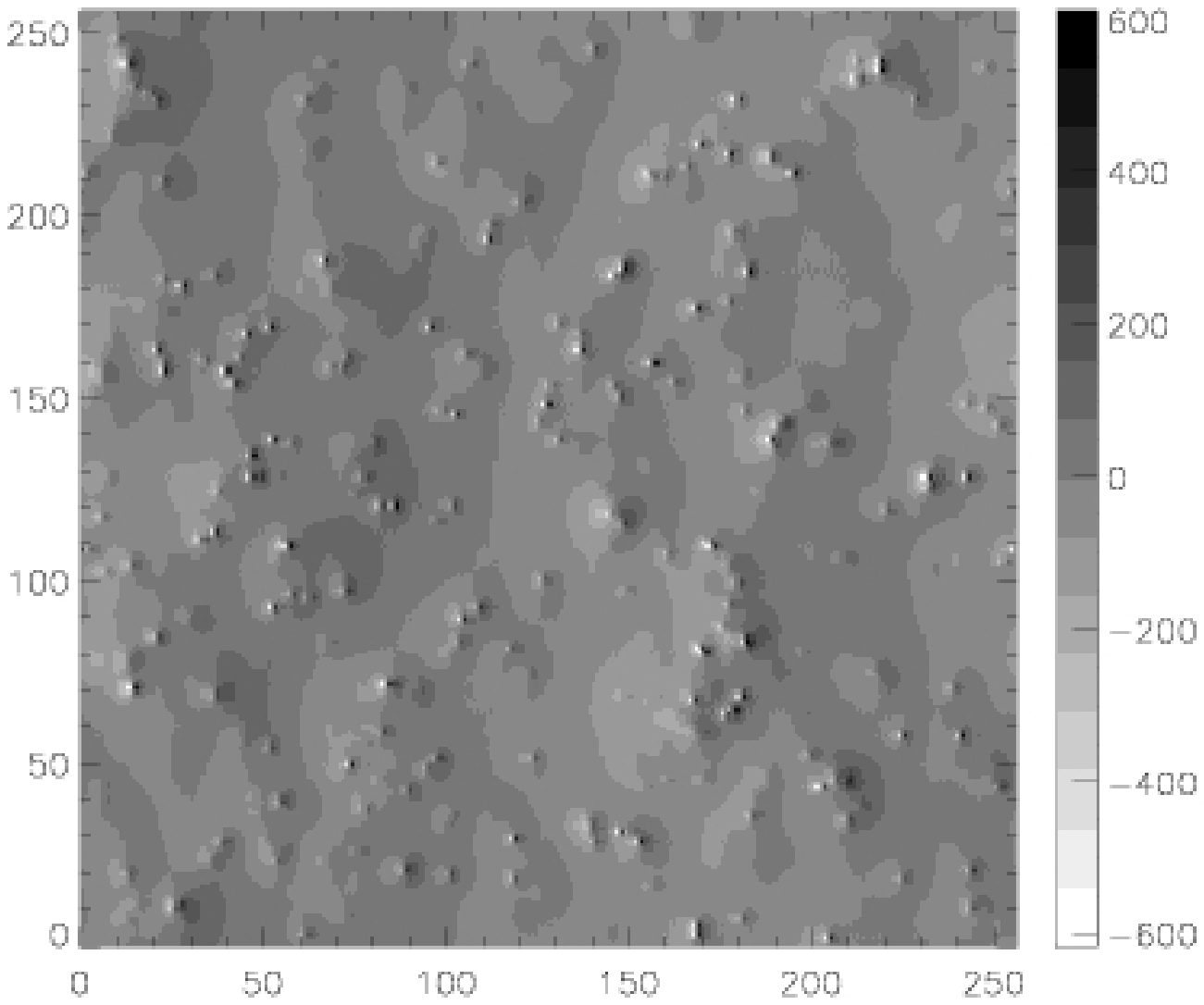}
\caption{The top panel shows velocity in z-direction for a slice in our simulation.
The middle panel shows
the reconstructed velocities of one slice of the simulation when we use all the dark matter
particles in the box.
The bottom panel shows the
reconstructed velocities of the same slice with Wiener filtering.
The velocities (only in z-direction) are pointing left and right of the map in both panel.}
\label{velcompare}
\end{figure}

We take this simulation and find its halos using a friends-of-friends (FOF) algorithm. After we identify the
halos, we use the Halo Occupation Distribution (HOD) from Bootes field~\cite{brown08} to populate the
halos with galaxies. The galaxies were selected in the Bootes field 
by enforcing uniform comoving number density of $\bar{n} = 10^{-3} (
mathrm{Mpc}/h)^{-3}$. We then only take these galaxies and compute the overdensity field. This allows us to assess the effects of incomplete sampling of the dark matter field by the galaxies.
 
To quantify the accuracy of the construction, we define a velocity reconstruction coefficient, $r(k)$, as:
\begin{equation}
\label{rkeq}
r(k) = \frac{\langle \mathbf{v}^{\mathrm{recon}}(k)^* \mathbf{v}^{\mathrm{sim}}(k) \rangle }{\langle \mathbf{v}^{\mathrm{recon}}(k)^* \mathbf{v}^{\mathrm{recon}}(k) \rangle },
\end{equation}
where $\mathbf{v}^\mathrm{recon}(k)$ are the reconstructed velocities, and $\mathbf{v}^{\mathrm{sim}}$ those found by the simulation itself. This is a simple way for us to gauge how good we can reconstruct velocities in the presence of both Poisson noise (from incomplete sampling of the dark matter field by galaxies) and breakdowns of Eq.~\ref{velocity} due to non-linear evolution.

We apply a Wiener filter to the density field before we reconstruct our velocity field, defined as follows:
\begin{equation}
\label{velnoisefilt}
W_v(k) = \frac{b^2 r^2(k) P(k)}{b^2 P(k)+\frac{1}{\bar{n}}},
\end{equation}
where $P(k)$ is the dark matter power spectrum, $b$ is a (possibly weighted) average bias for the survey, and $\bar{n}$ is the number density of the survey.

By comparing the top and bottom panels of Fig.~\ref{velcompare}, one can see the effect of this filter on 
the reconstruction; while small scale power in the velocity field is filtered out, the large scale velocities are traced reasonably well. The Weiner filter for generating the density field is simpler, since there is no additional ``reconstruction error'' as there is for the velocity:

\begin{equation}
\label{galnoisefilt}
W_{\mathrm{gal}}(k) = \frac{b^2 P^(k)}{b^2 P(k)+\frac{1}{\bar{n}}},
\end{equation}

%\begin{figure}
%\includegraphics[width=3.275in]{recon_from_part_vel_disbar_nofilter.ps}
%\includegraphics[width=3.275in]{recon_from_halo_with_filter.ps}
%\caption{The first panel shows the reconstructed velocities of one slice of the simulation when we use all the dark matter
%particles in the box. The second panel shows the reconstructed velocities of one slice of the simulation when we use
%''galaxies'' which are populated using a HOD on halos identified via friends-of-friends algorithm.}
%\label{fig:velcompare_npart_hod}
%\end{figure}

\begin{figure}
\includegraphics[angle=-90,width=3.275in]{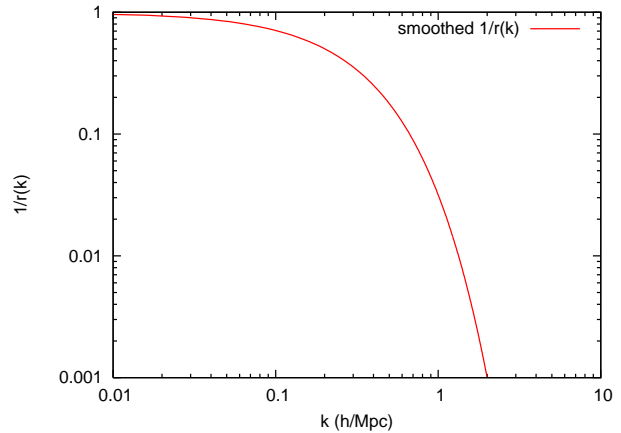}
\caption{Smoothed correlation coefficients ($1/r(k)$) between the reconstructed velocities and the real velocities are shown here.
At small k, the limited size of the simulation box provides us with insufficient information to reconstruct extremely
large scale velocities -- thus the drop off at small k. At large k, where non-linear effects kick in, the
linear theory we apply to reconstruct velocities is not valid anymore. We
however smoothed out the small k drop off, since this is an artifact of the 
limited simulation box size. We 
smoothed out the correlation coefficients at all scale so that we can apply a smooth
$r(k)$. However, for a large range of scales we have nearly perfect reconstruction.
}
\label{fig:rk}
\end{figure}

We have two main sources of noise: an inability to determine non-linear velocities fields, and Poisson noise from galaxies.
The effect of the latter can be seen in the bottom two panels of Fig.~\ref{velcompare}; the middle panel uses all the particles in the simulation and so is sensitive to non-linear breakdowns of Eq.~\ref{velocity}; the bottom panel uses only 
`galaxies` to reconstruct the velocities, thus is sensitive to both the non-linear breakdowns and also the Poisson noise from 
galaxies.

As we can see in a plot of $r(k)$ in Fig.~\ref{fig:rk}, the velocity field reconstructed
from the galaxy density field agrees well with both the velocity field reconstructed from the matter field and with the full velocity field of the simulation.
There is a significant range of $k$ in the linear regime that is well-reconstructed, while the correlation drops off at non-linear regime.

The studies of this section thus suggest that velocity reconstruction is a viable method even in the presence of the levels of noise
we might expect in a contemporary large-scale structure survey. We now turn our attention to making an analytic estimate of the kSZ signal.

\section{Estimating the kSZ Signal in $\ell$-space}
\label{kszcl}

Having validated our reconstruction method, we now turn our attention to predicting the dependence, on cosmological parameters, of the cross-correlation between a projected momentum field and the CMB.

\subsection{Theory}

The velocity field in the linear regime is a pure gradient field: the direction of the velocity is aligned with the
direction of its $k$-mode. Combining Eqs.~\ref{momentum} and~\ref{visibility} in the Limber approximation,
then, we can see that the source of the signal will be primarily from a modulation of this velocity field by
the density field. This ``second-order'' nature of the kSZ was realized early on~\cite{ov86}.

There is also the possibility of modulating the velocity field by the visibility function:
the so called ``velocity-Doppler'' effect that can be significant on large scales.
However this signal is drawn primarily from redshifts around reionization where the derivative of the
ionization fraction is large~\cite{ch00}; it does not correlate with a low-redshift survey.

To estimate the momentum-kSZ cross-correlation signal, we must first make assumptions about how well our
various observables trace the underlying dark matter and baryon fields. We take, as reasonable \emph{Ansatze} for the relationship
between baryon, dark matter and galaxy overdensities:
\begin{equation}
\label{baryonbias}
\delta_b(k) = f(k) \delta_m(k) 
\end{equation}
and
\begin{equation}
\label{galaxybias}
\delta_{\mathrm{gal}}(k) = b(k,z) \delta_m(k), 
\end{equation}
where $f(k)$ describes the difference in clustering power between baryons and dark matter -- we take the functional form of \cite{gh98} --
$b(k,z)$ is the scale and redshift dependent bias of galaxy overdensity field, and $\delta_m(k)$ is the
non-linear cold dark matter overdensity field. We take $\mathbf{v}_\mathrm{gal}$, our reconstructed velocity field discussed in the previous section, to be
\begin{equation}
\label{baryonest}
\mathbf{v}_{\mathrm{gal}}(\mathbf{k})=i\frac{d\ln{G}}{d\ln{a}}\frac{aH \delta_{\mathrm{gal}}(\mathbf{k})\mathbf{k}}{k^2},
\end{equation}
where, note, we do not divide out by the bias; this allows us to keep a handle on how bias estimates will affect our final result, and we similarly use $f(k)\delta_\mathrm{gal}$ as our estimate of baryon overdensities.

Once we apply the Wiener filters of Eqs.~\ref{velnoisefilt} and~\ref{galnoisefilt} to our estimates of the baryon velocity field, Eq.~\ref{baryonest}, and the baryon density field, we may Fourier transform into real space and project down to produce the reconstructed, projected momentum field, which we call $Q$.

We can then ask: how well does $Q$ correlate with $\Theta$? We here follow and elaborate on the analysis of Ref.~\cite{h00}. We know that $\mathbf{v}_\mathrm{gal}$ and $\mathbf{v}_\mathrm{b}$ are imperfect correlates, and assume that $r(k)$, defined in Eq.~\ref{rkeq}, captures all of these effects; this amounts to assuming that baryon velocities are governed entirely by gravity and so track the cold dark matter -- a reasonable assumption on the scales we consider. 
%{\tt Shirley, a good place to put in some references on this -- weren't you talking to Paul Bode about that assumption $i(k)=0$?}. 

We further assume that the effects of Poisson noise, non-linear velocity fields, and non-gravitational interactions are not correlated with the linear field, and that differences between dark matter and baryon flows on the relevant scales are negligible, so we may write,
\begin{eqnarray*}
\langle v_b(\mathbf{k}) v_\mathrm{gal}(\mathbf{k}^\prime) \rangle & = &  (2\pi)^3 \delta_D(\mathbf{k}-\mathbf{k}^\prime) \\
& & \times b(k,z) r(k) \left(\frac{d\ln{G}}{d\ln{a}}\frac{aH}{k}\right)^2 P(k),
\end{eqnarray*}
where $P(k)$ is the non-linear cold dark matter power spectrum; note that the correlation here depends also on bias.

We write the angular power spectrum of the cross-correlation as
\begin{equation}
\label{CLQT}
C_l^{\mathbf{Q}\Theta} = \frac{\pi^2}{2l^5}\int d\eta ~\eta^3 g(\eta)N(\eta)\left(\frac{\dot{G}(\eta)}{G(\eta)}\right)^2 I_{\mathbf{Q}\Theta}(l/\eta),
\end{equation}
where $\eta$ is comoving distance in units of Hubble distance as defined in Eq.\ref{comov}, $G(\eta)$ is the 
growth factor at $\eta$, and overdots are with respect to $\eta$. $I_{\mathbf{Q} \Theta}$ is the cross-power spectrum of the vorticity of the momentum field, 
which we can, after manipulation, write as
\begin{eqnarray}
\label{IQT}
I_{\mathbf{Q}\Theta}(k,z)&=& \int^\infty_0 dy_1\int^1_{-1} d\mu \frac{(1-\mu^2)k^6}{4\pi^4}  \nonumber \\
& & [b(k^\prime) f(k^\prime) P(k^\prime) W_{\mathrm{gal}}(k^\prime) \nonumber \\
& &  b(k^{\prime\prime}) r(k^{\prime\prime}) P(k^{\prime\prime}) W_v(k^{\prime\prime}) - \nonumber \\ 
& & \frac{y_1^2}{y_2^2}  b(k^\prime) r(k^\prime) P(k^\prime) W_{\mathrm{gal}}(k^\prime) \nonumber \\ 
& & b(k^{\prime\prime}) f(k^{\prime\prime}) P(k^{\prime\prime}) W_v(k^{\prime\prime})],
\end{eqnarray}
$k^\prime = ky_1$, $k^{\prime\prime} = ky_2$ and $y_2$ is equal to $\sqrt{1-2\mu y_1+y_1^2}$.

To compute the signal-to-noise, we will also need the auto-correlation of the template,
\begin{equation}
\label{CLQN}
C_l^{\mathbf{QQ}} = \frac{\pi^2}{2l^5}\int d\eta ~\eta^3 N(\eta)^2\left(\frac{\dot{G}(\eta)}{G(\eta)}\right)^2 I_{\mathbf{QQ}}(l/\eta).
\end{equation}
Since we Weiner filter both the density field and the velocity field in the construction of the 
template $\mathbf{Q}$, $I_{\mathbf{QQ}}$ is not identical to $I_{\mathbf{Q}\Theta}$; we find
\begin{eqnarray}
\label{IQQ}
I_{\mathbf{QQ}}(k,z)&=& \int^\infty_0 dy_1\int^1_{-1} d\mu \frac{(1-\mu^2)k^6}{4\pi^4} \nonumber \\ 
& & [b^2(k^\prime) P^{\prime}(k^\prime) W^2_{gal}(k^\prime) \nonumber \\
& &  b^2(k^{\prime\prime})  P^{\prime}(k^{\prime\prime}) W^2_v(k^{\prime\prime}) - \nonumber \\ 
& & \frac{y_1^2}{y_2^2}  b^2(k^\prime)  P^{\prime}(k^\prime) W_{\mathrm{gal}}(k^\prime) W_v(k^\prime) \nonumber \\
& & b^2(k^{\prime\prime})  P^{\prime}(k^{\prime\prime}) W_{\mathrm{gal}}(k^{\prime\prime}) W_v(k^{\prime\prime})]
\end{eqnarray}
We have here defined $P^{\prime}(k)$ as $P^{\prime}(k) + 1/b^2(k)\bar{n}$.

The ``science product'' consists both of a template $Q$, defined in this section, and an estimate, $C_l^{\mathbf{Q}\Theta,\mathrm{analytic}}$, of its cross-correlation amplitude with the CMB for a set of fiducial choices for the cosmological and galaxy survey parameters. The ratio, $R$ of this estimate to that actually measured with an experiment is
\begin{eqnarray}
\label{baryon}
R&=&  \frac{C_l^{\mathbf{Q} \Theta, \mathrm{observed}}}{C_l^{\mathbf{Q} \Theta,\mathrm{analytic}}} \nonumber \\ 
 &\propto &  b^2_\mathrm{eff} g_\mathrm{eff} 
\end{eqnarray}
where $b_{\mathrm{eff}}$ is an ``effective'' averaged bias, weighted by the various kernels $g(\eta)$ and $N(\eta)$, $g_\mathrm{eff}$ is the effective averaged visibility function within the survey volume that has been weighted
similarly as the bias. 
In the case that the fiducial parameters chosen are the true ones, $R$ is unity.

\subsection{Validating Estimates}

Since this is the first time this method has been constructed and applied, we check the validity of our calculation for $C_l^{\mathbf{Q}\Theta}$ through numerical simulations, in the following fashion:

\begin{enumerate}
\item We produce the ``real'' kSZ sky, $\Theta$, with an a N-body dark matter particles simulation box (as described earlier in Sec.~\ref{vsim}) from $z=0.428$ and assuming that the ionized gas traces the dark matter.
\item We produce the template, $Q$, with FOF (friends-of-friends) halos found from the same N-body and then use an HOD (from Ref.~\cite{brown08}) to populate the halos with their galaxies.
\item We project these two fields, and take their cross-correlation to find $C_l^{\mathbf{Q} \Theta, \mathrm{sim}}$.
\end{enumerate}

Note that the highly limited red-shift range of our simulations means that we will not reproduce an entire kSZ sky -- only that produced by baryons in a very limited range.

\begin{figure}
\includegraphics[width=2.275in,angle=-90]{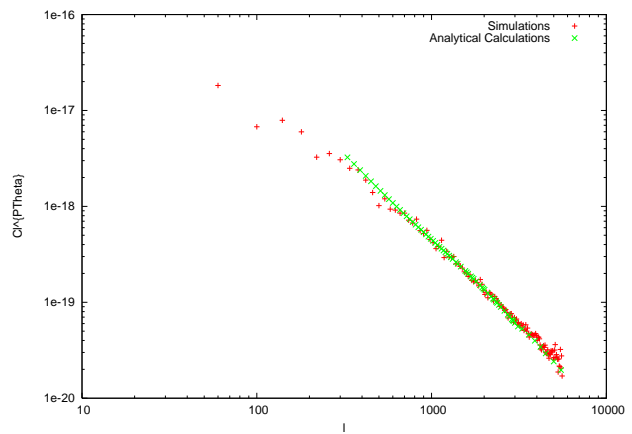}
\caption{We show the validity of our equations via comparison between the expected cross-correlation coefficient
 $C_l^{\mathbf{Q} \Theta, \mathrm{analytic}}$ and the calculated cross-correlation coefficient from simulations
 $C_l^{\mathbf{Q} \Theta, \mathrm{sim}}$.}
\label{fig:simcomp}
\end{figure}

Fig.~\ref{fig:simcomp} compares these two quantities; we find that at scales where there are a sufficient number of modes in the N-body to do our reconstruction reliably, the analytic calculations perform well and without appreciable bias. 
%The Pearson cross-correlation between the analytic and simulated $C_l^{\mathbf{Q}\Theta}$ is $0.9941$.

Given the expected signal-to-noise of upcoming measurements, then, the precision of our analytic formula for $I_{\mathbf{Q}\Theta}$ are more than sufficient for accurate determination of gas parameters.

\section{Estimating Signal-to-Noise for the kSZ cross-correlation signal}
\label{sn}
In this section, we evaluate whether current and upcoming surveys are capable of using the method developed in this paper 
to detect the kSZ-galaxy momentum template cross-correlation.  
Our method requires two very different types of observations: 
a sensitive high resolution CMB map (e.g., PLANCK, ACT or SPT) 
and a large-volume spectroscopic survey (e.g., SDSS, SDSS3, ADEPT).  
We estimate the expected signal/noise in the combined maps by computing:
\begin{displaymath}
\left(\frac{S}{N}\right)^2 = f_{\mathrm{sky}}\sum_l (2l+1)\frac{(C_l^{\mathbf{Q}\Theta})^2}{C_l^{\mathbf{QQ}}(C_l^{\mathrm{CMB}}+ C_l^{\mathrm{DET}})},
\end{displaymath}
where $f_{\mathrm{sky}}$ is the fraction of sky that is covered by both the spectroscopic survey and the high resolution CMB survey.
$C_l^{\mathbf{Q}\Theta}$ is as described in Eq.\ref{CLQT}, while $C_l^{\mathbf{Q} \mathbf{Q}}$ is simply the 2-point correlation function of 
$\mathbf{Q}$ (the momentum field) with
its noise. From the CMB side, we have $C_l^{\mathrm{CMB}}$ which is the CMB temperature anisotropies (which is noise here), and $C_l^{\mathrm{DET}}$ which is the detector noise from the CMB experiment.

We consider the following two CMB experiments specifically: 
\begin{enumerate}
\item Atacama Cosmology Telescope (hereafter ACT, see \footnote{{\tt http://www.physics.princeton.edu/act/}} for further information): we assume a $\theta_{\mathrm{FWHM}}$ of $1.4^{\prime}$ and a noise of $26$ $\mu$K~$\mathrm{arcmin}^{-2}$ with a survey area of $2000$ deg$^2$
based on a straw-man proposal.
\newline
\item PLANCK (see \footnote{{\tt http://www.rssd.esa.int/index.php?project=Planck}} for further information): we assume a $\theta_{\mathrm{FWHM}}$ of $7.1^\prime$ and a noise level of 
%$6$ $\mu$K per pixel 
$302$ $\mu$K~$\mathrm{arcmin}^{-2}$ for $75\%$ of the sky, since galactic foregrounds may prove to be hard to subtract from parts of the maps. 
\end{enumerate}

For large scale structure surveys, we consider two current surveys and a proposed mission concept: 
\begin{enumerate}
\item Sloan Digital Sky Survey (hereafter SDSS): we use the SDSS DR4 VAGC LSS sample \citep{blanton05} (which will 
be further described in Sec.~\ref{data}).
The bias of the main galaxy sample is found to be $\sim 1.2$ if we assume linear
bias (see Ref.~\cite{tegmark04}).  
For the purpose of signal-to-noise analysis, it is 
sufficient to assume linear bias, but we will discuss the effects of bias
later in Sec.~\ref{discussion}.
The bias of the spectroscopic LRGs is found to be $\sim 2$ (based on 
the powerspectrum analysis done in Ref.~\cite{eisenstein05}). 
We include bias as a free parameter throughout the theoretical calculation, with one exception. 
We only employ bias from other analysis when we generate filter functions ($W_{\mathrm{gal}}$ and $W_v$) as defined in Sec. ~\ref{vsim}.
\item Sloan Digital Sky Survey 3 (hereafter SDSS3, see \footnote{{\tt http://www.sdss3.org}}  for further information): We assume the 
availability of $10^6$  spectroscopic Luminous Red Galaxies 
over a quarter of the sky in the redshift range of $0.2 - 0.6$. 
Since SDSS3 plans to spectroscopically observe all of the photometric LRGs in SDSS, we 
use the redshift distribution and bias ($b \sim 2$) determined for the photometric LRGs in SDSS as 
described in Ref.~\cite{ho08}.
\item Advanced Dark Energy Physics Telescope (hereafter ADEPT, see \footnote{{\tt http://universe.nasa.gov/program/probes/adept.html}} for further information):
we assume the availability of $10^8$ galaxies
over $28600$ $\mathrm{deg}^2$  from $1<z<2$. 
We assume a bias of $1.5$ and a uniform distribution in comoving volume from $1<z<2$ here.
Since these galaxies are Lyman-alpha emitters, we take the bias from studies by Ref.~\cite{erb06} which suggests
a bias of $\sim 1.5$.
\end{enumerate}

For the $S/N$ calculation, we also assume an ionization fraction of 1 ($x_e=1$) and full hydrogen ionization.

\begin{table}
\begin{center}
\begin{tabular}{c|cc}
$S/N$      &     ACT (2000 $\mathrm{deg}^2$) &    PLANCK      \\
\hline
  SDSS-DR4  &    4.1             &  11.2 \\
  SDSS3       &  8.5              &  22.5 \\
  ADEPT     &    12.2             &  32.2

\end{tabular}
\end{center}
\caption{The signal-to-noise ratio for the detection of gas-momentum-kSZ correlation for different combinations of current and upcoming experiments.}
\label{sntable}
\end{table}

\section{Templates} 
\label{data}

%In order to look for missing baryons, we describe a method to 
%generate a template for the kinematic Sunyaev-Zel'dovich that can be
%used to detect the missing baryons.
%This momentum template is based on computing the product of a reconstructed velociy field
%with a galaxy field.  Thus, the combination of a galaxy redshift survey and
%a CMB survey can detect the ionized gas.
%In particular, we construct templates using data from Sloan Digital Sky Survey and 
%provide the relevant correlation coefficients for calculating the fraction of baryons
%in the Universe.
Finally, we illustrate our method by constructing a galaxy momentum template form the Sloan Digital Sky Survey. 
This template is a prediction for the ACT and PLANCK experimental specifications and this section demonstrates how to 
develop templates for other surveys.

\subsection{Data from Sloan Digital Sky Survey}

The Sloan Digital Sky Survey has acquired $ugriz$ CCD images of $10^4$ deg$^2$ of the high-latitude sky \citep{york00}. A dedicated 2.5m telescope
\citep{gunn98,gunn06} at Apache Point Observatory images the sky in photometric conditions \citep{hogg01} in five bands 
\citep{fukugita96,smith02} using a drift-scanning, mosaic CCD camera \citep{gunn98}. All the data processing are done by completely automated
pipelines, including astrometry, source identification, photometry \citep{lupton01,pier03}, calibration \citep{tucker06,padmanabhan07a}, spectroscopic
target selection \citep{eisenstein01,strauss02,richards02}, and spectroscopic fiber placement \citep{blanton03}. The SDSS is well underway, and has
produced seven major releases \citep{stoughton02,abazajian03,abazajian04,abazajian05,adel06,adel07a,adel07b}.

In this paper, we utilize mainly the SDSS DR4 VAGC  (\cite{blanton05}) sample and also the SDSS spectroscopic 
Luminous Red Galaxies (hereafter LRG, \cite{eisenstein01}), since these two samples are the largest spectroscopic samples publicly available and 
have near-uniform completeness over a large area of sky.
The spectroscopic LRG sample \cite{eisenstein01}, includes area beyond Data Release 4 (DR4). 
The total area coverage for this spectroscopic sample is 5154 square degrees, as available in the NYU Value Added Galaxy Catalog (VAGC \cite{blanton05}) at the time of the preparation of our project. 

\subsection{Constructing Templates}

We construct the templates by the following steps:
\begin{enumerate}
\item We select only sky areas that are covered by survey with completeness of over $85\%$,
so that the galaxy field nearly uniformly 
samples the underlying density field.
%A non-uniform sample also has more complex noise
%properties.
%This ensures us the galaxy sample we utilize samples the underlying density field with near uniformly, thus 
%we do not have to take into account of the noise that arises from the varying completeness of the sample.
\item We compute the overdensity field and embed the survey volume in a  
periodic box (of size $1762^3$ $(h^{-1}~\mathrm{Mpc})^3$, with cells of size $6.9^2$ $(h^{-1}~\mathrm{Mpc})^2$). We assume that the region without observations is at the mean density of the universe.
\item We compute the Wiener filtered density field and also the velocity field (via our six-step reconstruction method
as described in Sec.~\ref{vsection}). 
\item We compute the momentum field as it is projected along line of sight and calculate the 2D momentum field that would 
be used to cross-correlate with the appropriate CMB field to get the $C_l^{\mathbf{Q}\Theta}$. 
\end{enumerate}
\begin{figure}
\includegraphics[width=3.275in]{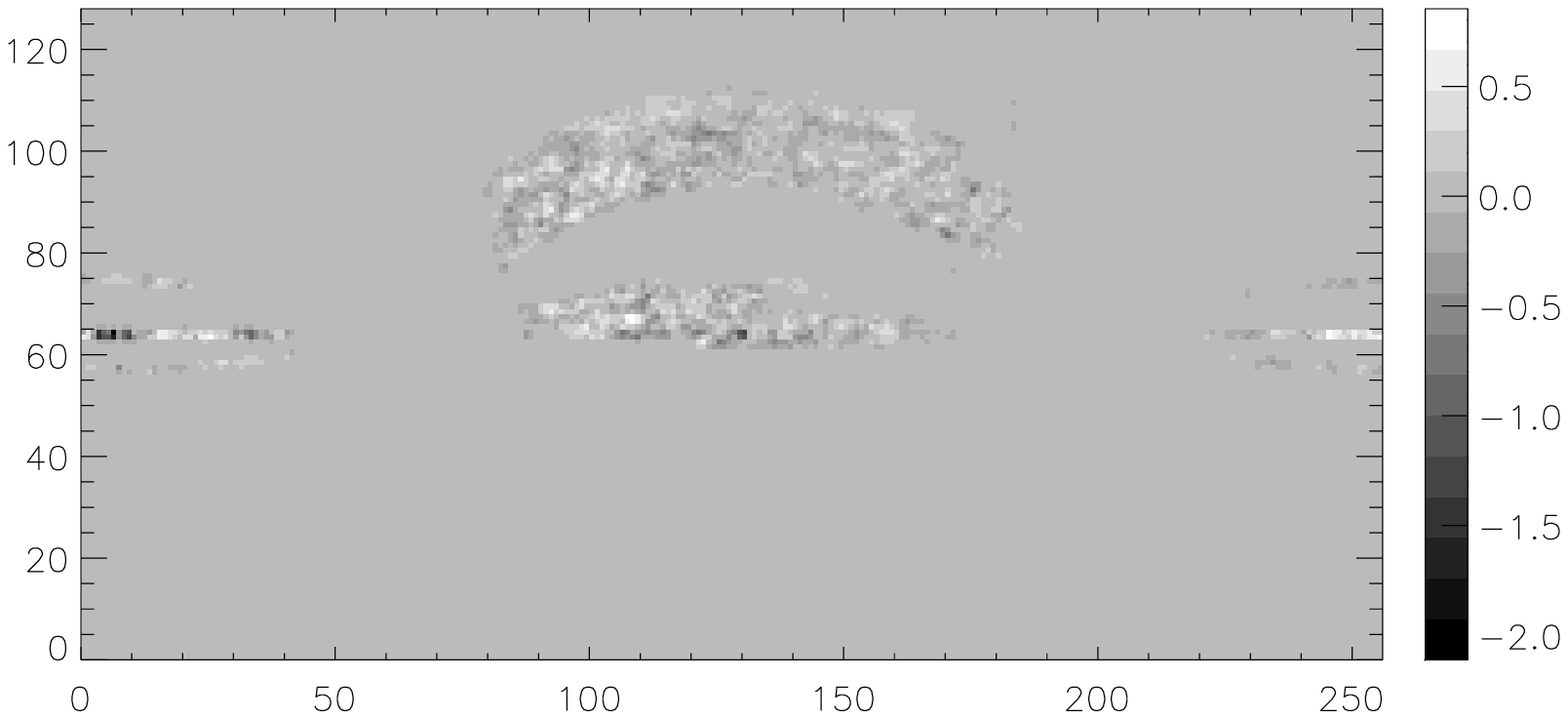}
\caption{Plotted is the 2D kSZ field made from SDSS DR4 Main galaxy momentum template.
The template predicts both hot and cold regions (due to the fact that this is a momentum field).
This implies that the template is likely to be nearly orthogonal to effects that are scale with
the galaxy or the matter density (such as integrated Sachs-Wolfe effect (\cite{sachs67}), Rees-Sciama effect \cite{rees68},
tSZ, lensing, dusty galaxy emission and to the galaxy foregrounds.
}
\label{temp1}
\end{figure}

We made the templates available in both ACT and various PLANCK resolutions. An update of the templates 
with full SDSS data-set will be made as data and time allow.
The templates are made available to public on our website \footnote{{\tt http://www.astro.princeton.edu/\~~shirley/SZ/SZ.html}}. 
Their respective $C_l^{\mathbf{Q}\Theta,\mathrm{analytic}}$ for the templates are also available as the filtering we use
in the paper affect the overall normalization of the signal. In general, it is simpler to 
compare the provided $C_l^{\mathbf{Q}\Theta,\mathrm{observed}}$ and the observed $C_l^{\mathbf{Q}\Theta}$ when 
one cross-correlates the templates with the CMB sky.

The two bias and gas-dependent parameters, $g_\mathrm{eff}$ (the weighted visibility function), and $b_\mathrm{eff}$ (the weighted bias) are complicated, but slowly varying, functions of the galaxy and gas fields; we take our fiducial model ($R$ of unity) to have constant, scale-independent bias $b$ of $1.2$ ($2$) for main galaxy (LRG) sample, and $\Omega_b $ of $0.0441$, $h$ of $0.71$ with $x_e$ of $1.0$. For reference, we plot the 2D kSZ field made from SDSS DR4 Main galaxy template here in Fig.\ref{temp1}. 
\begin{figure}
\includegraphics[width=3.5in]{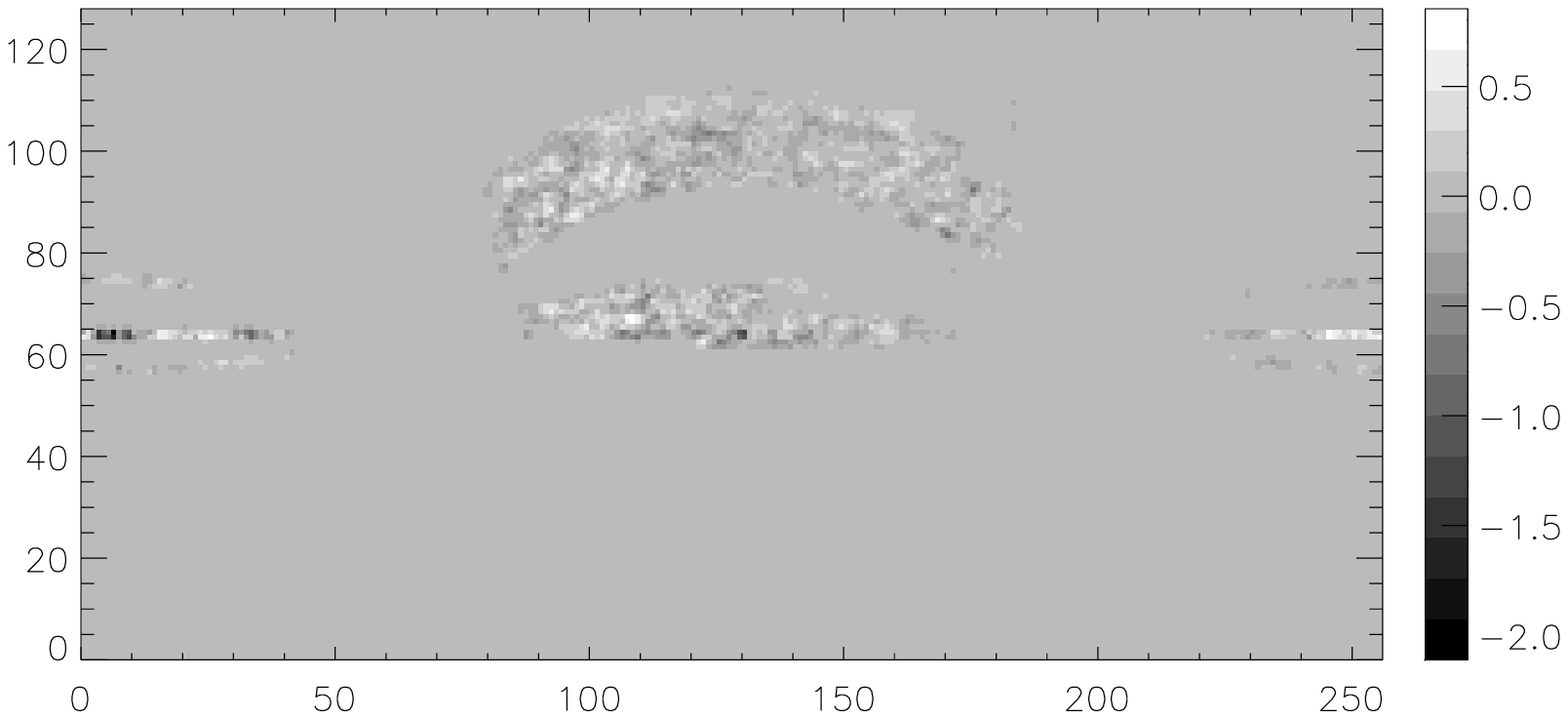}
\caption{The 2D kSZ template reconstructed from SDSS DR4 Value-added Galaxy Catalog (VAGC.)}
\label{temp1}
\end{figure}

\section{Discussion}
\label{discussion}
\subsection{Possible Caveats}

Our method deals well with a number of possible interfering signals; we discuss them here.
\begin{enumerate}
\item Thermal Sunyaev-Zel'dovich. The tSZ contributes to our estimates as a noise term, since tSZ signals are uncorrelated with the velocity fields; our 
momentum-kSZ correlation is not biased by the presence of tSZ. 
In principle, we can also use multifrequency data from CMB to separate the tSZ signal from kSZ 
as tSZ and kSZ have different spectral signatures. We can then reduce the noise contributed from 
tSZ. There is in general leakage of the tSZ signal into CMB maps due to uncertainties
in the detector frequency response function and due to relativistic tSZ effects (see Ref.~\cite{sehgal07}). Other contaminants that can not be picked out with frequency information, such as the Rees-Sciama (non-linear ISW) and lensing effects, are generally lower in amplitude; they, too, contribute to the noise, but again in an unbiased fashion.
\item Bias. 
Bias is treated as redshift and scale dependent throughout our theoretical calculations. 
However, as the ratio between the observed and theoretical $C_l$ are degenerate between 
the effective bias ($b^2_{\mathrm{eff}}$) and $g_\mathrm{eff}$, one can determine bias (with its redshift and scale dependence) but not being able to learn very much 
about the missing baryons. 

Since one can calculate bias from the power spectrum analysis of galaxy populations, we assume that 
bias can be determined fully in the large scale structure survey before one uses the survey to produce the 
gas-momentum field of the Universe.

For simplicity in our template production we 
assume scale and redshift independence for bias (which is appropriate for the scale and redshift we are working with)
in the production of the templates and take on values of bias which are predetermined via other efforts. 

In particular we use bias determined by \cite{tegmark04} in the construction of the filtering functions 
$W_{\mathrm{gal}}$ and $W_v$. 
In the calculation of signal-to-noise, we assume a bias for the galaxy population we (will) use to construct the template
for various experiments.
\item Dusty Galaxies. Star forming galaxies can be very luminous at submillimeter wavelength range, as suggested by 
Submillimeter Common User Bolometer Array (SCUBA) and other experiments (see \cite{blain02} for a review).
The emission is mainly from dust around the galaxies at approximately $\sim 10$ K, radiating mainly at sub-mm wavelength, 
but can also be significant at millimeter wavelengths.
Fortunately, the dusty galaxies do not correlate with the bulk flows of the Universe
and thus, by the same argument presented for tSZ, do not bias the measurement of the gas-momentum-kSZ correlation. 
\item Galactic Foregrounds. Galactic Foregrounds affect mostly the large-scale of the CMB and the galaxy observations
and since they do not correlate with the large-scale velocity field, we can 
safely assume that it would not bias our estimator, but would at most be adding to the noise term. 

We could also consider reducing the contribution of the noise from the galactic foregrounds in 
our cross correlation by limiting ourselves to certain multipoles where the galactic foregrounds 
are weak. We can gauge the effect of the galactic foregrounds by cross-correlating the momentum field with (for example)
dust extinction maps \cite{schlegel98} to find out the range of multipoles that are not 
affected by the galactic extinction.
\end{enumerate}

\subsection{Further applications}
%mention the baryon profiles in voids , around galaxies, etc.
We can use the gas-momentum-kSZ temperature correlation to constrain not only
the baryon content of selected regions of the Universe, but also 
the baryon profiles and evolution of its density.
We give examples below to describe how in theory we can apply 
the gas-momentum-kSZ temperature correlation to understand 
gas profiles and baryon evolution over redshift:
\begin{enumerate}
\item Gas profiles: 
Assuming that we know the bias and ionization fraction of the Universe in the same region 
of interest,
we can parameterize the electron density as a function of radius (from the center of the nearest galaxy or region under consideration) and 
generate the appropriate visibility functions. 
We can then construct momentum templates and 
maximize the cross-correlation coefficient between the electron density and 
the CMB to 
determine the best-fit parameters. This can be applied to finding gas profiles in voids or around different types of galaxies.
%For example, if we like to understand the amount of gas inside voids: 
%we would parameterize $f_{ionized}$ as an increasing function of $r$, where 
%$r$ here is defined from the center of voids, which can be found by 
%various methods such as triangular tessellation (see further details in Ref.~\cite{conroy05}, and 
%then make the momentum templates (with void density field and reconstructed velocities 
%that we calculate from galaxy density field). Then, we can maximize 
%the cross-correlation as described earlier and determine $f_{ionized}(r)$ for 
%void regions. 
%Furthermore, we can look at the different gas profiles around for example, red
%and blue galaxies and investigate the relationship between color of galaxies
%and their gas profiles, lending us hints to the underlying principles of galaxy formation.
\item Baryon content at different redshifts. 
With ever increasing size and volume coverage of galaxy redshift surveys, 
we can create momentum templates based on galaxy surveys binned at different 
redshifts. We can then trace the evolution of electron density through a large 
redshift range.
\end{enumerate}

\section{Conclusion}

We have presented a new method of 
generating a template for the kinetic Sunyaev-Zel'dovich that can be
used to detect the missing baryons.
This momentum template is based on computing the product of a reconstructed velocity field
with a galaxy field.  Since the kinetic Sunyaev-Zel'dovich effect 
is a line of sight integral of electron momenta (modulo 
constants such as Thompson scattering cross-section), the combination of a galaxy redshift survey and
a CMB survey can detect the ionized gas.

Unlike other techniques that look for hot gas or metals, this approach directly detects the
electrons in the IGM through their signature on the CMB.
%Many different methods for detecting low-redshift baryons have been proposed, but while some methods promise a higher signal-to-noise, few are as theoretically ``clean'' as a detection of the kSZ. 
Since the kSZ is produced simply by Thompson scattering from free electrons, 
there is no need for detailed knowledge about the metallicity of the medium and its evolution, nor for an understanding of potentially out-of-equilibrium level populations. 

The kSZ amplitude is the product of the gas density and its velocity, 
and the latter, as we have shown, can be well-modeled by linear theory, providing an independent check, on different scales and relying on different physics, from baryon surveys that employ tracers that scale as the square of density, or that are more concentrated in the center of the most massive virialized objects.

Our studies in this paper find that a CMB survey, with sufficient resolution to 
push past the ``damping tail'' of the primordial fluctuations but not necessarily covering the entire sky, partnered with 
an overlapping galaxy survey with sufficient number density to find structures at these angular 
scales, can provide an unambiguous detection of the baryons. 
%In particular, a combination of ACT and a survey such as SDSS3, will be able to do so in the near future.

We have estimate the expected signal-to-noise for detecting the galaxy-momentum kSZ cross-correlation
for a few different combinations of current and upcoming experiments.
The estimated signal-to-noise for detecting the galaxy-momentum kSZ cross-correlation
is 4.1, 8.5, 12.2 for ACT (with survey area of 2000 $\mathrm{deg}^2$) with SDSS-DR4, SDSS3 and ADEPT.
The estimated signal-to-noise  for PLANCK with SDSS-DR4, SDSS3 and ADEPT is 11.2, 22.5 and 32.2.
The proposed estimator provides an exciting avenue into understanding the ionized gas in the Universe in
the near future.

We have produced momentum templates from Sloan Digital Sky Survey, so that 
the scientific community make take arcminute scale CMB data from current and upcoming
experiments such as ACT, SPT and PLANCK and determine not only the gas fraction of the Universe, but also 
the distribution and evolution of the gas fractions 
and gas profiles around different regions in the sky.

\begin{acknowledgments}
We thank Paul Bode for his help in simulations, Martin White, Jerry Ostriker, Jo Dunkley and Joe Hennawi for insightful
comments. 
Shirley Ho and David Spergel's work on this project has been partially supported by the NASA WMAP project, 
NASA grant NNX08AH30G and NSF grant 0707731.

Funding for the SDSS and SDSS-II has been provided by the Alfred P. Sloan Foundation, the Participating Institutions, the National Science Foundation,
the U.S. Department of Energy, the National Aeronautics and Space Administration, the Japanese Monbukagakusho, the Max Planck Society, and the Higher
Education Funding Council for England. The SDSS Web Site is http://www.sdss.org/.

The SDSS is managed by the Astrophysical Research Consortium for the Participating Institutions. The Participating Institutions are the American Museum
of Natural History, Astrophysical Institute Potsdam, University of Basel, University of Cambridge, Case Western Reserve University, University of
Chicago, Drexel University, Fermilab, the Institute for Advanced Study, the Japan Participation Group, Johns Hopkins University, the Joint Institute
for Nuclear Astrophysics, the Kavli Institute for Particle Astrophysics and Cosmology, the Korean Scientist Group, the Chinese Academy of Sciences
(LAMOST), Los Alamos National Laboratory, the Max-Planck-Institute for Astronomy (MPIA), the Max-Planck-Institute for Astrophysics (MPA), New Mexico
State University, Ohio State University, University of Pittsburgh, University of Portsmouth, Princeton University, the United States Naval Observatory,
and the University of Washington.

\end{acknowledgments}

\bibliographystyle{apsrev}
\bibliography{ms}

\end{document}